\documentstyle[12pt]{article}
\input epsf
\newcommand{\beq}{\begin{eqnarray}}   
\newcommand{\eeq}{\end{eqnarray}}

\newcommand{\gsim}{\lower.7ex\hbox{$
\;\stackrel{\textstyle>}{\sim}\;$}}
\newcommand{\lsim}{\lower.7ex\hbox{$
\;\stackrel{\textstyle<}{\sim}\;$}}

\input epsf
\newcommand{\grpicture}[1]
{
    \begin{center}
        \epsfxsize=300pt
        \epsfysize=200pt
        \vspace{-5mm}
        \parbox{\epsfxsize}{\epsffile{#1.eps}}
        \vspace{5mm}
    \end{center}
}

\begin{document}
\begin{titlepage}

\begin{flushright}
ITEP-TH-2/98

\end{flushright}
\vspace*{3cm}

\begin{center}
{\Large \bf 6 + 1 Vacua in Supersymmetric QCD with $G_2$ Gauge Group}
\vspace{2cm}

{\Large  A.V. Smilga} \\

\vspace{0.8cm}

{\it ITEP, B. Cheremushkinskaya 25, Moscow 117218, Russia}\\

\end{center}

\vspace*{2cm}

\begin{abstract}
We consider ${\cal N} = 1$ 
 supersymmetric $QCD$ based on the $G_2$  gauge 
group and involving 3 chiral matter 7--plets $S_\alpha^i$. 
In that case, the gauge symmetry is broken completely {\it and}
the instanton--generated superpotential on the classical moduli
space is present. If the theory involves the Yukawa term 
$\lambda f^{\alpha\beta\gamma} S^1_\alpha S^2_\beta S^3_\gamma$, there are 
{\it six} chirally asymmetric
vacua. In the limit $\lambda \to 0$, two of the vacua run away to
infinity and only 4 asymmetric vacuum states are left. Besides, a 
chirally symmetric state is always present.
We consider also an $O(7)$ model with 4 chiral multiplets in spinor
representation. In that case, there are 4 extra ``virtual vacua'' dwelling
at infinity of the moduli space. In a non-renormalizable theory with a
quartic term in the superpotential, they show up at finite moduli values.

\end{abstract}

\end{titlepage}

\section{Introduction}

The dynamics of the ${\cal N} = 1$ supersymmetric QCD has been 
 studied by theorists since the
beginning of the eighties \cite{brmog}. The theories with unitary gauge 
groups attracted
a special attention. In the most simple from ideological viewpoint case when
the number of flavours is $N_f = N_c -1$, the gauge symmetry is broken
completely and the theory involves a discrete set of vacuum states.
The existence of $N_c \equiv N$ such  states with nonvanishing  value of
the gluino condensate $<{\rm Tr} \ \lambda^2>$ associated with the spontaneous
breaking of a discrete symmetry $Z_{2N} \to Z_2$ has been known for a long
time.
 It was noted recently \cite{Kovner} that on top of $N$ chirally
asymmetric vacua  also a chirally symmetric vacuum with the 
zero value of the condensate exists. 

The vacuum structure of the theory displays itself in a straightforward way
in the framework of the effective lagrangian due to Taylor, Veneziano, and
Yankielowicz (TVY) \cite{TVY}. The lagrangian is written for the
colorless composite fields (moduli)
 \beq
  \label{normN}
\Phi^3 = \frac 1{16\pi^2} {\rm Tr}\ W^2, \ \ {\cal M}_{ij} =\ 2 
\tilde{S}_iS_{j},
\ \ \ \ \ \ \ i,j = 1,\ldots, N-1
  \eeq
where $W$ is the chiral gauge superfield and $\tilde{S}^\alpha_i, S_{j\alpha}$
are the matter chiral multiplets in the antifundamental and fundamental
representations of the gauge group, respectively. The TVY effective lagrangian
presents
a Wess--Zumino model with the superpotential
  \beq
  \label{TVY}
{\cal W} =  \Phi^3 \left[ \ln \frac{\Phi^3 {\rm det} {\cal M}}
{\Lambda_{SQCD}^{2N + 1}} \ -\ 1 \right] - \frac{m}{2} {\rm Tr}\ {\cal M} 
\eeq  
($m$ is the common mass for all matter fields).
 The corresponding potential for the lowest scalar components $\phi, \ 
 \mu_{ij}$ of the superfields $\Phi, \ {\cal M}_{ij}$ has $N+1$ degenerate
minima. One of them is chirally symmetric $\phi =  \mu_{ij} = 0$, and
there are also N chirally asymmetric vacuum states.  

The latter are clearly seen in the weak--coupling limit $m \ll 
\Lambda_{\rm SQCD}$ where we can integrate out the heavy field $\Phi$
associated with the gauge degrees of freedom (i.e. to freeze down
$\Phi^3 {\rm det} {\cal M} \equiv \Lambda_{\rm SQCD}^{2N+1}$
 \footnote{It {\it is} the step where the chirally symmetric state is lost.
 The latter appears if freezing down $\Phi \equiv 0 $. The valleys
 $\Phi^3 {\rm det} {\cal M}  \equiv \Lambda_{\rm SQCD}^{2N+1}$ and $\Phi
\equiv 0 $ in
 the TVY lagrangian are separated by an ostensibly high but still penetrable
 energy barrier \cite{KSS,SV,SUN}.}
) to obtain

  \beq
  \label{Higgs}
  {\cal W} = -  \frac{\Lambda_{\rm SQCD}^{2N + 1}}
  {{\rm det} {\cal M}} - \frac{m}{2} {\rm Tr}\ {\cal M} 
  \eeq
The effective lagrangian with the superpotential (\ref{Higgs})
is a true Born--Oppenheimer lagrangian for the light fields and has a better
status than the TVY lagrangian (\ref{TVY}) where light and heavy degrees
of freedom are not so nicely separated. The existence of $N$ chirally
 asymmetric vacua as follows from Eq.(\ref{Higgs}) is a {\it theorem} of
 the supersymmetric QCD while the presence of the chirally symmetric
 vacuum following from Eq.(\ref{TVY}) is a {\it conjecture}. We believe
 it is true, but the question is still under discussion
  \cite{symdisc}.

The existence of the 
discrete set of vacua implies the presence of the domain walls interpolating 
between them \cite{Dvali,KSS,Chib,Witwal}. The extensive numerical study of 
these walls
\cite{KSS,SV,SUN} displayed their rich non-trivial structure. 
A certain kind of the
walls (those connecting different chirally asymmetric vacua) exists only for
small enough masses. Sometimes the walls are BPS saturated and
sometimes they are not, there are ``wallsome sphalerons'', etc. 

The theories with orthogonal and exceptional groups also attracted a 
considerable attention. They are interesting, in particular, because simple 
arguments
leading to the estimate $I_W = {\rm <rank\ of\ the\ group>} + 1$ for the 
number
of vacuum states \cite{Witssb} which work well for the unitary and simplectic
groups ( the chirally symmetric vacuum  appears only in the infinite
volume limit \cite{KSS,SV}
 and its presence does not invalidate the finite volume
calculation of Ref. \cite{Witssb}) fail in
this case. Also, orthogonal and exceptional groups do not 
involve a sufficiently rich center subgroup and
the so called toron field configurations \cite{toron}
\footnote{Such configurations appear formally when a theory with a unitary
gauge group is defined on a large 4--dimensional torus. They provide an 
essential contribution in the Euclidean path integral when  the size of the 
box is small. The question of whether  such configurations play an essential 
role in
the limit of large boxes is not yet totally clear. See \cite{QCD2,KSS,SV} 
for a recent discussion.} 
are absent.

The previous studies of the theories with exotic groups displayed the
following dynamical picture:

 The number of (chirally asymmetric) vacuum states in the pure
  supersymmetric gluodynamics is equal to the Dynkin index of the group
  $T(G)$ defined as ${\rm Tr} \{T^a T^b\} = T(G) \delta^{ab}$ where
  $T^a$ are the generators in the adjoint representation. This is best seen
  by noting that, for a general group, instantons involve $2T(G)$ gluino
  zero modes. Supersymmetric Ward identities + an explicit instanton 
  calculation require that the chiral correlator
  $<T\{\lambda_\alpha^a \lambda^{a\alpha}(x_1) \ldots 
  \lambda_\alpha^a \lambda^{a\alpha}(x_{T(G)})\}>$ 
  (here $\alpha = 1,2$ is the Weyl spinor index) is a non-zero $x_i$ --
  independent constant. By cluster decomposition, that implies the existence 
  of the states where the vacuum expectation value $<\lambda_\alpha^a
  \lambda^{\alpha a}>$ is nonzero. $T(G)$ values for the phase of the gluino
  condensate are allowed \cite{brmog,Witssb}. 

  For unitary and symplectic groups, $T(G) = r + 1$ in accordance with 
  the original Witten's counting \cite{Witssb}. For higher orthogonal
  groups [starting from $O(7)$] and for exceptional groups, $T(G)
  > r + 1$. A ``traditional'' explanation for this mismatch was that
  the method how the estimate $r+1$ for the number of states was obtained
  is not quite rigourous. In was based on the Born--Oppenheimer treatment
  of the theory on a small spatial torus with topologically trivial
  boundary conditions. But however small the torus is, the parameter
  of the Born--Oppenheimer expansion is not small everywhere.
  An explicit calculation for supersymmetric QED displayed large corrections
  to the lowest order Born--Oppenheimer hamiltonian in some region
  of the moduli space \cite{jaQED}. Such corrections could in principle
  invalidate the vacua counting. Quite recently, however, Witten found a 
  bug in his original reasoning \cite{Witnew}: it turned out that, for 
  complicated groups, the moduli
  space has two disconnected components. The second component is responsible
  for some extra vacuum states so that their total number is just $T(G)$ !
  This was checked explicitly for  orthogonal groups.
  
   $T(G)$ supersymmetric vacua are seen also in the supersymmetric
  QCD involving light chiral matter multiplets \cite{Cordes}. However, up to 
  now, only the theories involving the tree--level mass term were considered.
  
The purpose of this note is to demonstrate that, in the general case
when the tree--level lagrangian involves also Yukawa couplings, the number
of vacua may be larger. In particular, the $G_2$ theory with 3 chiral
7--plets involves generally {\it six} chirally asymmetric vacua.

\section{Vacuum Structure in $G_2$ theory}
\setcounter{equation}0
$G_2$ theory is defined as a subgroup of $O(7)$ leaving invariant
the combination $f^{\alpha\beta\gamma} p_\alpha q_\beta r_\gamma$ where
$p_\alpha,\ q_\beta$, and $r_\gamma$ are three arbitrary 7-vectors and
 $f^{\alpha\beta\gamma}$  is a certain antisymmetric tensor. One particular
choice for $f^{\alpha\beta\gamma}$ is 
  \beq
  \label{f}
  f^{165} = f^{341} = f^{523} = f^{271} = f^{673} = f^{475} = f^{246} = 1
  \eeq
  and all other nonzero components are restored by antisymmetry. The form
  (\ref{f}) can be mnemonized by drawing the triangle diagram as in Fig. 1.
  $f^{\alpha\beta\gamma}$  is nonzero only for the indices lying on the
 same line, with the arrows indicating the order of indices when
  $f^{\alpha\beta\gamma}$ is positive.

\unitlength=1mm

    \begin{figure}
   \grpicture{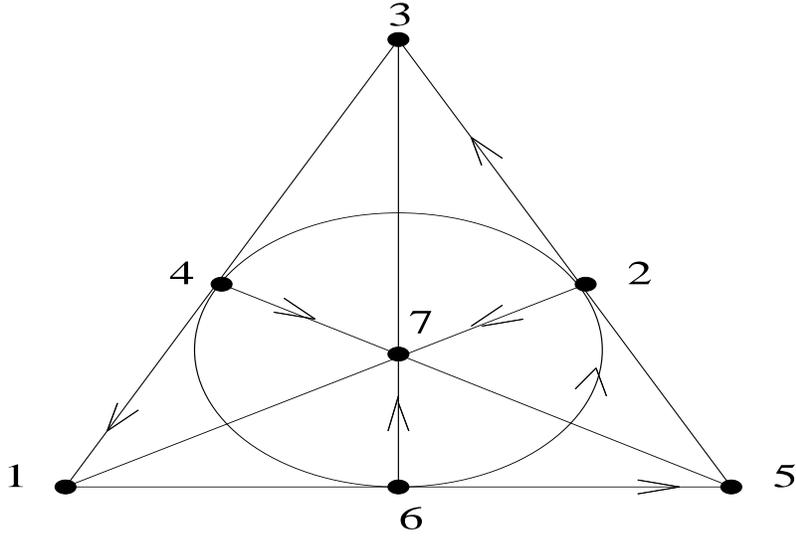}
\caption{Triangle septagram for $f^{\alpha\beta\gamma}$.}
    \end{figure}

  Another way to describe the $G_2$ group is to think of it as of a subgroup
  of $O(7)$ leaving invariant a real 8--component 7D spinor $\eta$.
 14 generators out of 21 generators of $O(7)$ are left. 7 others act
  on the spinor $\eta$ non--trivially. The rank of $G_2$ is $r=2$, one
  unit less than the rank of $O(7)$. In this
  construction, the tensor  $f^{\alpha\beta\gamma}$ is defined as
  \beq
  \label{feta}
   f^{\alpha\beta\gamma} = \eta^T \Gamma^\alpha \Gamma^\beta \Gamma^\gamma
    \eta
 \eeq
 where $\eta^T\eta = 1$ and $\Gamma^\alpha$ are 7D Euclidean 
 (real and antisymmetric)
 $\Gamma$--matrices. The particular form (\ref{f}) is obtained from 
 (\ref{feta}) if choosing
 \beq
 \label{Gammas}
 \Gamma^1 &=& i\sigma^2 \otimes \sigma^2 \otimes \sigma^2,
 \Gamma^2 =  -i \otimes \sigma^1 \otimes \sigma^2, 
  \Gamma^3 = -i \otimes \sigma^3 \otimes \sigma^2,
\Gamma^4 = i \sigma^1 \otimes \sigma^2 \otimes 1, \nonumber \\
 \Gamma^5 &=&  -i \sigma^3 \otimes \sigma^2 \otimes 1,
 \Gamma^6 = -i \sigma^2 \otimes 1 \otimes \sigma^1,
\Gamma^7 =  -i\sigma^2 \otimes 1 \otimes \sigma^3,
\eeq
and
  \beq
  \label{eta0}
  \eta \ =\ \left( \begin{array}{c} 1\\0 \end{array} \right) \otimes
  \left( \begin{array}{c} 1\\0 \end{array} \right) \otimes
  \left( \begin{array}{c} 1\\0 \end{array} \right) 
  \eeq
The quantities $v_\alpha = \Gamma_\alpha \eta$ form a real 7--plet which
  is the fundamental representation of $G_2$.

The theory we want to discuss involves the gauge supermultiplet $V^a$ and
3 chiral 7--plets $S^i_\alpha$, $i = 1,2,3$. Note that though the
representation {\bf 7} of $G_2$ is real, the chiral fields $S^i_\alpha$,
and in particular their lowest scalar components $s^i_\alpha$ are complex.

As we will shortly see, the ground state of the theory corresponds to
nonzero values of $<s^i_\alpha>$ which break down completely the $G_2$
gauge group. Indeed, a nonzero vacuum expectation value of one of the fields
$<s^1_\alpha> = v_0 \delta_{\alpha 7}$ breaks $G_2$ down to $SU(3)$. 
After such a breaking, two other 7--plets   are decomposed as
${\bf 7} = {\bf 1} + {\bf 3} + {\bf \bar 3}$. 
Explicitly: $s^{2,3}_7$ form the singlets, and the triplets $r_\alpha,
t_\alpha$ and antitriplets $\tilde{r}^\alpha, \tilde{t}^\alpha$ are
  \beq
  \label{tripl}
  r_\alpha, t_\alpha = \frac 1{\sqrt{2}}
\left( \begin{array}{c} s^{2,3}_1 + i s^{2,3}_2 \\
  s^{2,3}_3 + i s^{2,3}_6 \\ s^{2,3}_5 + i s^{2,3}_4 \end{array} \right),\ 
  \tilde{r}^\alpha, \tilde{t}^\alpha = \frac 1 {\sqrt{2}} \left(
  s^{2,3}_1 - i s^{2,3}_2, s^{2,3}_3 - i s^{2,3}_6, s^{2,3}_5 - i s^{2,3}_4
  \right)
  \eeq  
As is well known \cite{brmog},
two triplets and two antitriplets which are left is a minimal necessary set
of matter fields to break the remaining $SU(3)$ group completely. 

The moduli space for this theory is constructed in the same way as for unitary
groups. If the tree--level superpotential is zero (i.e. if the mass of
the matter fields and their Yukawa couplings are zero), the classical 
vacuum energy vanishes whenever the $D$ -- term vanishes:
 \beq
 \label{D}
 D^a \ =\ \sum_i \bar s_i T^a s_i \ =\ 0
 \eeq
 This imposes 14 conditions on $3\times 7 = 21$ complex or 42 real parameters
 describing a general scalar field configuration. Further 14 parameters are
 absorbed by group rotations. Thus, the classical moduli space involves
 14 real or 7 complex bosonic parameters promoted by supersymmetry up to
 7 chiral superfields. They can be conveniently chosen as
  \beq
  \label{moduli}
  {\cal M}_{ij} = S_{i\alpha} S_{j\alpha},\ \ \ B = \frac 16 \epsilon_{ijk} 
  f^{\alpha\beta\gamma} S^i_\alpha S^j_\beta S^k_\gamma
  \eeq

  Quantum effects result, however, in this theory in a nontrivial 
  instanton--induced superpotential on the moduli space. Holomorphy and
  symmetry considerations dictate
  \beq
  \label{WholG2}
  {\cal W}^{\rm inst} ({\cal M}_{ij}, B) = -\frac {\Lambda^9_{G_2}}
  {{\rm det} {\cal M}} f\left( \frac {B^2}
   {{\rm det}  {\cal M}} \right)
   \eeq
   where $\Lambda_{G_2}$ is the characterestic scale for the 
   glueball (and gluino-ball) states in the $G_2$ supersymmetric gauge theory.
   The particular form of the function $f$ is $f(x) = 1/(1-x)$
   \cite{WG2}. The easiest way to see it is to require that, after the 
 breaking $G_2 \to SU(3)$ induced by the vacuum expectation value of one of
 the fields $<s^1_\alpha> = v_0 \delta_{\alpha 7}$, the
  instanton--induced superpotential 
 (\ref{WholG2}) would coincide with the one in the $SU(3)$ theory with
 2 chiral triplets and 2 antitriplets:
   \beq
  \label{WSU3}
  {\cal W}^{\rm inst}_{G_2}  \to {\cal W}^{\rm inst}_{SU(3)} =
   -\frac {\Lambda^7_{SU(3)}}
  {(\tilde{R}^\alpha R_\alpha)  (\tilde{T}^\alpha T_\alpha) -
  (\tilde{R}^\alpha T_\alpha) (\tilde{T}^\alpha R_\alpha) } 
   \eeq
[ $\Lambda_{SU(3)}^7 \equiv \Lambda_{G_2}^9/(4v_0^2)$ ].
 
  The theory with the superpotential
    \beq
    \label{WG2}
    {\cal W}^{\rm inst}  = -\frac {\Lambda^9_{G_2}}
    {{\rm det}  {\cal M} - B^2} 
    \eeq
    does not have a ground state at all --- the energy is positive
    at any finite value of the moduli. When adding to Eq.(\ref{WG2}) the
    tree--level superpotential, a discrete number of vacuum states appear.
    To find how many, we have to study the full effective theory with the
    superpotential 
     \beq
    \label{WfullG2}
    {\cal W}  = -\frac {1}
    {2({\rm det}  {\cal M} - B^2)}  - \frac m2 {\rm Tr} \ {\cal M}
    - \lambda B
    \eeq
  (from now on, everything will be measured in the units of $\Lambda_{G_2}$,
  and the factor 2 is put downstairs in the first term for convenience).
Basically, we have to find the points in the moduli space where all the
F--terms vanish:
 \beq
  \label{Fvac}
  \partial {\cal W}/\partial({\rm moduli}) = 0
  \eeq

Let   us see first what happens when $\lambda = 0$. Assuming 
${\cal M}_{ij} = V_0^2 \delta_{ij}, B=0$ and substituting it in the
 superpotential (\ref{WfullG2}), the condition (\ref{Fvac}) is reduced to
   \beq
   \label{Flam0}
   m(v_0^2)^4 = 1
   \eeq
   The equation (\ref{Flam0}) for the moduli $v_0^2$ has 4 roots corresponding
  to 4 chirally asymmetric vacuum states. This result is well known 
 \cite{Cordes}.

 Let us now study the general case $\lambda \neq 0$. The previous Ansatz
 with $B=0$ does not go through the equations (\ref{Fvac}) anymore. We can
 still assume ${\cal M}_{ij} = V^2_0 \delta_{ij}$, but have to allow for a
 nonzero $B$. 

 Note first that the equations (\ref{Fvac}) are written somewhat symbolically
 and should be handled with some care. Effectively, one can write them down
 in our case as
  \beq
  \label{natmod}
  \partial {\cal W}/\partial \mu_{ij} = 0,\ \ \ \partial {\cal W}/
 \partial b = 0 
  \eeq
  All the solutions of the equation system (\ref{natmod}) describe true
  supersymmetric vacua.
However, some other choice of variables (defining e.g. the superfield of
canonical dimension 1 $C = B^{1/3}$ and solving the equation 
$\partial{\cal W}/\partial c = 0$) could lead to extra fake solutions.
 To write the equations {\it quite} correctly, we have to take accurately 
 into account the {\it metrics} on the moduli space , in other words to
 choose the variables so that the kinetic term in the effective lagrangian
 would have a standard quadratic form.

 The latter is induced by the kinetic term of the matter fields in the
 fundamental theory:
  \beq
  \label{kin}
  {\cal L}^{\rm kin}_{\rm matt} = \int d^4\theta \sum_i \bar S^i S^i
  \eeq
  To proceed quite generally, one should resolve the constraint (\ref{D}),
  choose a group orientation,  express $S^i_\alpha$ via 7 complex moduli
  and  substitute these expressions in Eq.(\ref{kin}). This is a complicated
  technical problem which has not yet been solved. Fortunately, we do not
 need this. It suffices to act in the framework of the Ansatz
  \beq
  \label{Ans}
  {\cal M}_{ij} = V_0^2 \delta_{ij},\ \ \ B \neq 0
  \eeq
  In that case, a simple convenient parametrization for the matter
 superfields can be chosen:
   \beq
  \label{param}
  S^1_\alpha &=& V_0 \delta_{\alpha 7} \nonumber \\
  S^2_7 &=& 0, \ R_\alpha = \frac 1{{2}} \left( \begin{array}{c}
  iV_1\\ V_2 \\ 0 \end{array} \right), \ \tilde{R}^\alpha =
  \frac 1{{2}} \left( -iV_1, V_2, 0 \right) \nonumber \\
  S^3_7 &=& 0, \ T_\alpha = \frac 1{{2}} \left( \begin{array}{c}
  V_1\\ iV_2 \\ 0 \end{array} \right), \ \tilde{T}^\alpha =
  \frac 1{{2}} \left( V_1, -iV_2, 0 \right) ,
  \eeq
where $R_\alpha, T_\alpha$ and $\tilde{R}^\alpha, \tilde{T}^\alpha$
are defined in Eq. (\ref{tripl}). The parametrization (\ref{param})
corresponds to the vanishing D--terms as it should and implies the
following values of the moduli:
${\cal M}_{11} = V_0^2, \ {\cal M}_{22} = {\cal M}_{33} = (V_1^2 + V_2^2)/2,
\ {\cal M}_{i\neq j} = 0$. Many other parametrizations with
 $D^a = 0$ and the same values of the moduli are possible, but they all 
 are obtained from Eq.(\ref{param}) by colour and flavour rotations.

 If we want to keep the condition (\ref{Ans}), the constraint $2V_0^2 =
 V_1^2 + V_2^2$ should be imposed. But it is not really necessary: one
 can as well keep three free parameters $V_0, V_1, V_2$, and the condition
 $2v_0^2 = v_1^2 + v_2^2$ will be automatically satisfied at the vacuum
 points where all the F--terms (\ref{Fvac}) vanish. With the parametrization
 (\ref{param}), the kinetic term (\ref{kin}) has a simple form
   \beq
  \label{kinmod}
  {\cal L}^{\rm kin}_{\rm matt} = \int d^4\theta ( \bar V_0 V_0
  + \bar V_1 V_1 + \bar V_2 V_2)
  \eeq
  That means that the correct form of the equation (\ref{Fvac}) for the 
  vacuum states is just $\partial {\cal W}/\partial v_0 =  
  \partial {\cal W}/\partial v_1 = \partial {\cal W}/\partial v_2  =0$.
  Expressing the superpotential (\ref{WfullG2}) via $V_0, V_1, V_2$:
  \beq
  \label{WVi}
  {\cal W} = - \frac 1{2V_0^2 V_1^2 V_2^2} - \frac m2 (V_0^2 + V_1^2
  + V_2^2) - \frac \lambda 2 V_0(V_2^2 - V_1^2) ,
  \eeq
  the equations (\ref{Fvac}) acquire a form
   \beq
    \label{eqvac}
    mv_0 + \frac \lambda 2 (v^2_2 - v_1^2) = \frac 1{v_0^3 v_1^2 v_2^2} 
    \nonumber \\
    mv_1 - \lambda v_0 v_1 = \frac 1{v_0^2 v_1^3 v_2^2}  \nonumber \\
 mv_2 + \lambda v_0 v_2 = \frac 1{v_0^2 v_1^2 v_2^3}  
   \eeq
   This equation system has 12 solutions, the six pairs of them differing
   only by the sign of $v_0$ (and of $v_1^2 - v_2^2$) and corresponding
   to the same values of the moduli $v_0^2,\ b = v_0(v_2^2 - v_1^2)/2$.
  The relation $b = -\lambda v_0^4/m$ holds while $u = v_0^2$ satisfies
  the equation
   \beq
   \label{equ}
   mu^4 \left(1 - \frac {\lambda^2}{m^2} u \right)^2 = 1
   \eeq
   The equation (\ref{equ}) has 6 roots as was announced. 

   Let us see what happens in the limit of small $\lambda$, the smallness
   being characterized by a dimensionless parameter 
    \beq
    \label{kappa}
   | \kappa| = \left| \frac{\lambda^2}{m^{9/4}} \right| 
    \equiv \left| \lambda^2 (\Lambda/m)^{9/4} \right| \ll 1
    \eeq
    (do not forget that $\lambda$ and $m$ are generally complex).
    Then 4 of the roots are very close to the known solutions of the equation
    (\ref{Flam0}). but on top of this, thee are two extra roots at {\it large}
values of $u$: $ u = m^2/\lambda^2 \pm \lambda^2/m^{5/2}$. In the limit
$\lambda \to 0$, these new vacua run away at infinity of the moduli space
and decouple.

The latter statement can be attributed a precise meaning. Different vacua
are separated by the domain walls --- planar field configurations
 interpolating     between one vacuum on the left and another vacuum
 on the right. The domain walls have a surface energy density $\epsilon$.
 The value of $\epsilon$ is the measure of the height of the barrier
 separating different vacua. It turns out
\cite{BPS},\cite{Dvali},\cite{Chib}
 that the wall energy density satisfies a strict lower bound
  \beq
  \label{bound}
  \epsilon \geq 2|{\cal W}_1 - {\cal W}_2 |
  \eeq
 where ${\cal W}_{1,2}$ are the values of the superpotential at the vacua 1,2
 between which the wall interpolates. The bound (\ref{bound}) has the
 same nature as the celebrated BPS lower bound for the mass of the magnetic
 monopole. It often happens that an actual supersymmetric domain
 wall is BPS saturated, i.e. its energy density is given by the bound
 (\ref{bound}). But sometimes it is not so \cite{SV,SUN}. Only a detailed 
 numerical  study can answer the question what kind of  domain
 walls exist in this model, whether they are BPS saturated or not, and how
 that depends on the value of the parameter $\kappa$. 

Something can be said, however. First, when $\kappa = 0$ and only 4 vacua
are left, the domain walls between  them are BPS--saturated and have the
energy density $\epsilon = 2\sqrt{2} |m|^{3/4} $ for the walls connecting the
``adjacent'' vacua with, say, $u = m^{-1/4}$ and $u = i m^{-1/4}$  and
 $\epsilon = 4|m|^{3/4}$ for the walls connecting the
``opposite'' vacua with $u = \pm m^{-1/4}$ or with  $u = \pm i m^{-1/4}$.
The point is that the effective Higgs lagrangian is exactly the same here 
as for the $SU(4)$ theory with 3 chiral quartets and 3 antiquartets
whose domain walls were studied in  Ref.\cite{SUN}.
When $\kappa$ is nonzero and small so that a couple of new vacua at large
values of $u$ appear, the values of the superpotential (\ref{WVi})
at these vacua are also large $\approx m^3/2\lambda^2$. The bound 
(\ref{bound})
dictates that the energy density of a wall connecting  an ``old'' and a 
``new'' vacua is larger than $|m^3/\lambda^2|$ and tends to infinity in the 
limit $\lambda \to 0$.

When we increase $|\kappa|$, the new vacua move in from infinity and, at
$|\kappa| \sim 1$, the values of $u_{\rm vac}$ for the vacua of both types
 are roughly the same. It is interesting that, at $\kappa = \frac 2{3\sqrt{3}}
 \sqrt[4]{1} \approx .385 \sqrt[4]{1}$, two of the vacua (an ``old'' one and
  a ``new'' one) become
 degenerate. At $|\kappa| \gg 1$
 \footnote{To assure that $u_{\rm vac}$ are still large (compared to 
 $\Lambda$) so that we are still in the Higgs phase and the light and heavy
 degrees of freedom are separated, we should keep $|\lambda/m^{3/4}| \ll 1$.
 But if $m$ is small enough, this condition can be fulfilled at arbitrary
 large $|\kappa|$.},
 6 vacua find themselves at the vertices of a perfect hexagon on the complex
 $u$ -- plane. The complex roots of Eq.(\ref{equ}) in the units of $m^{-1/4}$
 for 3 illustrative values of $\kappa$ are displayed in Fig. 2.

\unitlength=.6mm

 \begin{figure}
 \begin{picture}(220,100)
 \put(-20,50){\line(1,0){90}}
 \put(0,25){\line(0,1){55}}
\put(-9.3,50){\circle*{1.5}}
 \put(-.7,40.2){\circle*{1.5}}
 \put(-.7,59.8){\circle*{1.5}}
 \put(11,50){\circle*{1.5}} 
 \put(60.8,50){\circle*{1.5}}
 \put(64,50){\circle*{1.5}}
\put(20,50){\line(0,1){1.5}}
 \put(20,53){{\scriptsize 2}} 
 \put(40,50){\line(0,1){1.5}}
 \put(40,53){{\scriptsize 4}} 
 \put(0,70){\line(1,0){1.5}}
 \put(3,70){{\scriptsize 2}} 
 \put(0,30){\line(1,0){1.5}}
 \put(3,30){{\scriptsize -2}}
\put(20,10){{\small a)}}
 \put(48.6,78.6){{\small u}} 
 \put(50,80){\circle{5}}

 
 \put(90,50){\line(1,0){60}}
 \put(105,25){\line(0,1){55}} 

 \put(96.4,50){\circle*{1.5}}
 \put(103.5,40.7){\circle*{1.5}}
 \put(103.5,59.3){\circle*{1.5}}
 \put(122.3,48.5){$\times$} 
 \put(134,50){\circle*{1.5}}
 
 \put(125,50){\line(0,1){1.5}}
 \put(125,53){{\scriptsize 2}} 
 \put(145,50){\line(0,1){1.5}}
 \put(145,53){{\scriptsize 4}} 
 \put(105,70){\line(1,0){1.5}}
 \put(108,70){{\scriptsize 2}} 
\put(105,30){\line(1,0){1.5}}
 \put(108,30){{\scriptsize -2}}
 
 \put(115,10){{\small b)}}
 \put(138.6,78.6){{\small u}} 
 \put(140,80){\circle{5}}

 \put(170,50){\line(1,0){50}}
 \put(195,25){\line(0,1){50}}
 
\put(173.8,50){\circle*{1.5}}
 \put(184.6,31.4){\circle*{1.5}}
 \put(184.6,68.6){\circle*{1.5}}
 \put(206.1,31.4){\circle*{1.5}} 
 \put(206.1,68.6){\circle*{1.5}}
 \put(216.8,50){\circle*{1.5}}
 
\put(210,50){\line(0,1){1.5}}
 \put(210,53){{\scriptsize 0.2}} 
 \put(170,50){\line(0,1){1.5}}
 \put(170,53){{\scriptsize -0.2}} 
 \put(195,70){\line(1,0){1.5}}
 \put(198,70){{\scriptsize 0.2}} 
 \put(195,30){\line(1,0){1.5}}
 \put(198,30){{\scriptsize -0.2}}
 
\put(195,10){{\small c)}}
 \put(213.6,78.6){{\small u}}  
 \put(215,80){\circle{5}}
 \end{picture}

 \caption{ Solutions of the equation $u^4(1 - \kappa u)^2 = 0$
 for {\it a)} $\kappa = .16$, {\it b)} $\kappa = .385$, 
 and {\it c)} $\kappa = 100$. The cross marks out the double degenerate
root at $\kappa = .385$.}

 \end{figure}
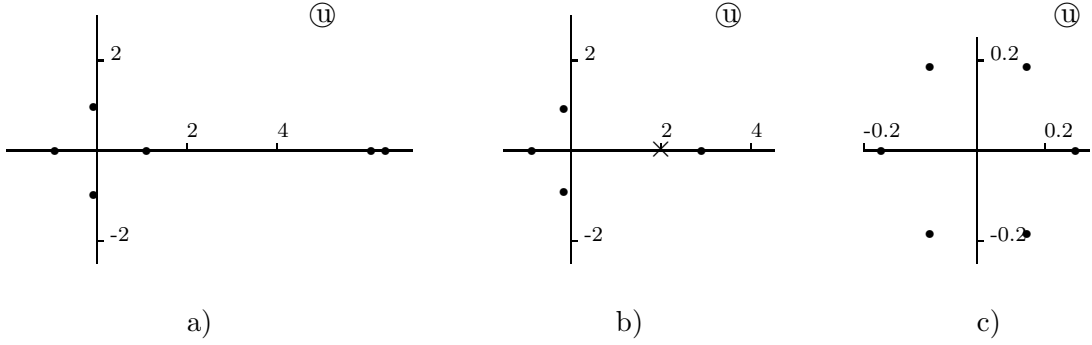

 \begin{figure}
 \begin{picture}(220,100)
 \put(-20,50){\line(1,0){60}}
 \put(15,25){\line(0,1){50}}
 
\put(35.7,50){\circle*{1.5}}
 \put(14.2,70.1){\circle*{1.5}}
 \put(14.2,29.9){\circle*{1.5}}
 \put(-4.1,50){\circle*{1.5}} 
 \put(-17.8,50){\circle*{1.5}}
 \put(-14.7,50){\circle*{1.5}}
 
\put(35,50){\line(0,1){1.5}}
 \put(35,53){{\scriptsize 2}} 
 \put(-5,50){\line(0,1){1.5}}
 \put(-5,53){{\scriptsize -2}} 
 \put(15,70){\line(1,0){1.5}}
 \put(18,70){{\scriptsize 2}} 
 \put(15,30){\line(1,0){1.5}}
 \put(18,30){{\scriptsize -2}}
 
\put(15,0){{\small a) $\kappa$ = .16}}
 \put(33.4,78.6){{\small w}} 
 \put(35,80){\circle{6}}

 
 \put(80,50){\line(1,0){50}}
 \put(105,25){\line(0,1){50}} 

 \put(116.7,50){\circle*{1.5}}
 \put(103.2,70.3){\circle*{1.5}}
 \put(103.2,29.7){\circle*{1.5}}
 \put(95.6,50){\circle*{1.5}} 
 \put(87.7,48.5){$\times$}
 
 \put(125,50){\line(0,1){1.5}}
 \put(125,53){{\scriptsize 2}} 
 \put(85,50){\line(0,1){1.5}}
 \put(85,53){{\scriptsize -2}} 
 \put(105,70){\line(1,0){1.5}}
 \put(108,70){{\scriptsize 2}} 
\put(105,30){\line(1,0){1.5}}
 \put(108,30){{\scriptsize -2}}
 
 \put(105,0){{\small b) $\kappa$ = .385}}
 \put(123.4,78.6){{\small w}} 
 \put(125,80){\circle{6}}

 \put(160,50){\line(1,0){70}}
 \put(190,15){\line(0,1){70}}
 
 \put(225.9,50){\circle*{1.5}}
 \put(173.1,81.1){\circle*{1.5}}
 \put(173.1,18.9){\circle*{1.5}}
 \put(172.1,79.2){\circle*{1.5}} 
 \put(172.1,20.8){\circle*{1.5}}
 \put(223.7,50){\circle*{1.5}}
 
\put(210,50){\line(0,1){1.5}}
 \put(210,53){{\scriptsize 4}} 
 \put(170,50){\line(0,1){1.5}}
 \put(170,53){{\scriptsize -4}} 
 \put(190,70){\line(1,0){1.5}}
 \put(193,70){{\scriptsize 4}} 
 \put(190,30){\line(1,0){1.5}}
 \put(193,30){{\scriptsize -4}}
 
\put(190,0){{\small c)$\kappa$ = 100}}
 \put(228.4,78.6){{\small w}}  
 \put(230,80){\circle{6}}
 \end{picture}

 \caption{ Values of the superpotential ${\cal W}_{\rm vac}$.}

 \end{figure}
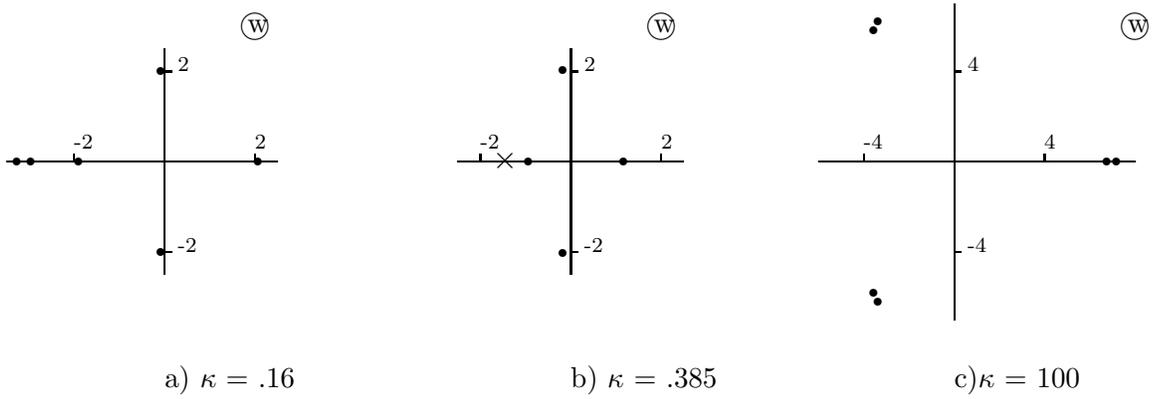 

 The values of the superpotential ${\cal W}(u_{\rm vac}, b_{\rm vac})$
 for the same values of $\kappa$ in the units of $m^{3/4}$ :
 ${\cal W}_{\rm vac} = -2u_{\rm vac} + (3/2) \kappa u_{\rm vac}^2$ 
 are shown in Fig. 3. We see that, for large
 $\kappa$, 6 vacua are clustered in 3 pairs (each pair corresponding to
 the opposite vertices of the hexagon in Fig. 2c). The values of the
  superpotential for  two vacua of the same pair are close and hence the
  energy barrier between them is small. Indeed, for very small masses and
  fixed $\lambda$, one can neglect the mass term in the superpotential
  in which case the vacuum solutions occur at $v_0^2 = 0$ while the value
  of $b$ is fixed at $b = \sqrt[3]{-1/\lambda}$. The splitting of vacua 
  inside the pair is the effect of higher order in $1/\kappa$.
 
All 6 vacua are chirally asymmetric, i.e. the gluino condensate
$< \lambda^a \lambda^a>$ is formed. The latter is fixed by the Konishi
identity \cite{Konishi}
 \beq
 \label{Konishi}
 \sum_i T(R_i) < \lambda^a \lambda^a> = - 16\pi^2 \left< Q_i \frac {\partial 
 {\cal W}^{\rm tree}}{\partial Q_i} \right>
 \eeq
 where $T(R_i)$ are Dynkin indices of the group representations where the 
 matter fields $Q_i$ lie (in our case, $Q_i \equiv
 S^i_\alpha$, T({\bf 7}) = 1 and $\sum_i T(R_i) = 3$). We have
  \beq
  \label{cond}
 < \lambda^a \lambda^a>_i = \frac{16\pi^2}3 [3mv_0^2 + 3\lambda b] = 
  16\pi^2 m u_i \left( 1 - \frac {\lambda^2}{m^2} u_i \right)
  \eeq
  where $u_i$ is the $i$-th root of the equation (\ref{equ}). For small
  $|\kappa|$, $< \lambda^a \lambda^a> \sim m^{3/4}$ for old vacua while
  for the new vacua the gluino condensate is suppressed 
  $ < \lambda^a \lambda^a> \sim \lambda^2 m^{-3/2} \sim \kappa m^{3/4}$.
  When $\lambda \to 0$ it vanishes altogether. However, these new
  states have nothing to do with the chirally symmetric state of
 Ref.\cite{Kovner}. The latter is realized at zero values of all moduli
 (including also the moduli describing  gauge degrees of freedom) 
and there are BPS domain walls with finite energy density connecting
the true chirally symmetric vacuum with chirally asymmetric ones.
 The states we have found have large vacuum expectation values for all moduli,
a large superpotential ${\cal W}(u_{\rm vac}, b_{\rm vac})$
\footnote{In more simple models with unitary groups and also in the theory
in hand with $\lambda = 0$, the full effective superpotential coincides
with the gluino condensate up to a numerical factor common for all
vacua. But it is not so in general case. } ,
and decouple in the limit $\lambda \to 0$.
 
\section{$0(7)$ theory}
\setcounter{equation}0

The next in complexity theory where $T(G) \neq r+1$ is the theory based
on the $O(7)$ gauge group. The rank of $O(7)$ is $r = 3$ while $T[O(7)]
=5$. Consider the ${\cal N} = 1$ supersymmetric gauge model involving
4 chiral matter multiplets $S^i_\alpha$ in spinor representation. 
In this theory, the gauge symmetry is completely broken. Indeed, v.e.v.
of one of the spinor fields $<S^1_\alpha> \sim \eta$  break $O(7)$
down to $G_2$ after which 3
other spinors are decomposed as ${\bf 8} \to {\bf 1} + {\bf 7}$
(explicitly: $S^2_\alpha = (P_0 + P_\mu \Gamma_\mu) \eta$ etc ). 
Three 7--plets break down the remaining $G_2$
completely as was discussed in the previous section.

Originally, the theory involved 4$\times$8 = 32  complex degrees of
freedom in the
matter sector. Imposing the conditions $D^a = 0$ and fixing the gauge
eliminates 21 degrees of freedom so that the classical moduli space
involves 11 complex parameters. 10 of them can be presented as 
${\cal M}_{ij} = S_i^T S_j$. To construct the eleventh one, note
that the tensor product of two $O(7)$ spinors involves a scalar, a vector,
an antisymmetric tensor with two indices and an antisymmetric tensor with
3 indices: {\bf 8}$\times${\bf 8} = {\bf 1} + {\bf 7} + {\bf 21} + {\bf 35}.
Multiplying it over by itself, we obtain 4 different scalars in the tensor
product ${\bf 8} \times {\bf 8} \times {\bf 8} \times{\bf 8}$.
Three of these scalars can be written as $(S_{1}^T S_2)
 (S_{3}^T
S_4)$, $(S_{1}^T S_3) (S_{2}^T S_4)$, and 
$(S_{1}^T S_4) (S_{2}^T S_3)$ while the fourth one presents
a non-trivial quartic group invariant not reducible to the products
${\cal M}_{12} {\cal M}_{34}$ etc. Its explicit form is

 \beq
 \label{invO7}
 H = \frac 1{24} \epsilon^{ijkl}   
 (S^{T}_i \Gamma_\mu S_j) (S^{T}_k \Gamma_\mu S_l)
  \eeq
  The invariant (\ref{invO7}) is antisymmetric in flavour and colour
  indices.

  Like in the previous case, one can show that the theory involves an
 instanton--generated superpotential. Holomorphy and
 symmetry considerations dictate 
   \beq
   \label{WholO7}
   {\cal W}^{\rm inst}_{O(7)} = - \frac{\Lambda^{11}_{O(7)}}
   {{\rm det} {\cal M} } f\left( \frac {H^2}
   {{\rm det}  {\cal M}} \right)
   \eeq 
The function $f$ is fixed by the requirement that after the breaking induced
by the vacuum expectation value of one of the spinor fields $<S^1_\alpha>
= v_0 \eta$, the superpotential (\ref{WholO7} ) is reduced to the 
superpotential (\ref{WG2}) in the $G_2$ theory. Noting that 
${\rm det} {\cal M}_{O(7)} \to -v_0^2 {\rm det} {\cal M}_{G_2}$ and
$H \to - v_0 B$, we obtain 
    \beq
    \label{WO7}
    {\cal W}^{\rm inst}  = -\frac {\Lambda^{11}_{O(7)}}
    {{\rm det}  {\cal M} + H^2} 
    \eeq
The main difference with the $G_2$ theory is that we are not able here 
to write down a Yukawa term in the superpotential on the tree level.
No group invariant can be constructed out of just 3 spinors.

Let us relax, however, for a moment the requirement that the theory is
renormalizable. Then one could add the term $\sigma H$ with a
dimensionful coupling $\sigma$ in the superpotential. The full
superpotential would have the form quite analogous to Eq. (\ref{WfullG2}):
    \beq
    \label{WfullO7}
    {\cal W}  = -\frac {1}
    {2({\rm det}  {\cal M} + H^2)}  - \frac m2 {\rm Tr} \ {\cal M}
    + \sigma H
    \eeq
($\Lambda \equiv 1$). Proceeding in the same way as in the $G_2$ case, we
search for the solutions of the equation (\ref{Fvac}) in the form
$\mu_{ij} = v_0^2 \delta_{ij}; h \neq 0$. One can show that the
equation (\ref{Fvac}) written in correct variables implies
$h = -\sigma u^3/m$ where $u = v_0^2$ satisfies the equation
 \beq
 \label{equO}
 mu^5 \left(1 - \frac{\sigma^2}{m^2}  u^2 \right)^2 = 1
 \eeq
 The equation (\ref{equO}) has nine solutions. For small $|\sigma|$ 
 ($|\sigma^2/m^{12/5}| \ll 1$), 5 of them are the conventional ones 
 $u = \sqrt[5]
 {1/m}$ while 4 others are new solutions with large values of $u \sim 
 \pm m/\sigma$. They are separated from the conventional vacua by a high
 BPS energy barrier $\epsilon \sim m^2/\sigma$. If allowing
 $|\sigma^2/m^{12/5}| \sim 1$, $u_{\rm vac}$ for all 9 vacua are of the same
 order. At $\sigma^2/m^{12/5} = 5/9 (2/3)^{8/5} \sqrt[5]{1}$
 (the analog of the point $\lambda^2/m^2 =  2/({3\sqrt{3}}) \sqrt[4]{1}$
 for the $G_2$ theory), two of the vacua become degenerate. For large
 $\sigma/m^{6/5}$, $u_{\rm vac}$ are placed in the vertices of a perfect
 9-gon while ${\cal W}(u_{\rm vac}, h_{\rm vac})$ are groupped in 3 clusters 
 with 3 vacua each.

 However, we cannot really consider a non-renormalizable theory. A theory
 with $\sigma = 0$ and involving only the mass term in the superpotential
 has just 5 vacua. 4 extra ``virtual'' vacua dwell at infinity of the 
 moduli space and are separated from the conventional states by an
 infinitely high barrier.

 \section{Chirally symmetric phase.}
\setcounter{equation}0

The presence of the chirally symmetric phase is not seen in the framework
of the effective Higgs lagrangian (\ref{WfullG2}). To see it, we have
to study the full TVY effective lagrangian involving also the superfield
$\Phi^3 =  W^a W^a /(32\pi^2)$ describing the gauge degrees of freedom.
The analog of Eq.(\ref{TVY}) for the $G_2$ theory is
   \beq
  \label{TVYG2}
{\cal W} =  \Phi^3 \left[ \ln \frac{\Phi^3 ({\rm det} {\cal M} - B^2)}
{\Lambda^9} \ -\ 1 \right] - \frac{m}{2} {\rm Tr}\ {\cal M}
- \lambda B 
\eeq  
 The superpotential (\ref{TVYG2}) is rigidly fixed by the requirements:
 {\it i)} the chiral and conformal anomalies of the underlying gauge
 theory are reproduced correctly, {\it ii)} when $m = \lambda = 0$,
 the effective theory is symmetric under the transformations
$\theta \to \theta e^{i\beta}, {\cal M}_{ij} \to {\cal M}_{ij} e^{-2i\beta/3},
B \to B e^{-i\beta}, \Phi \to \Phi e^{2i\beta/3}$ induced by 
 the anomaly--free chiral transformation in the underlying theory
 $\theta \to \theta e^{i\beta}, W^a \to W^a e^{i\beta},\ S^i_\alpha
 \to S^i_\alpha e^{-i\beta/3}$,
 and {\it iii)} when integrating out the heavy field $\Phi$ by
 freezing down $\Phi^3 ({\rm det} {\cal M} - B^2)  \equiv 1$ 
 , the effective Higgs lagrangian (\ref{WfullG2}) is reproduced.

 If choosing the  symmetric Ansatz ${\cal M}_{ij} = V_0^2 \delta_{ij}$,
 adding to the superpotential (\ref{TVYG2}) kinetic terms (As was
 mentioned before, the kinetic terms for the moduli ${\cal M}_{ij}, B$
 are induced by the kinetic terms of the matter fields in the original
 theory. The kinetic term for the moduli $\Phi$ is not known and we just
 choose it in the simplest possible form $\int \bar \Phi \Phi d^4\theta$.),
 and solving the equations (\ref{Fvac}), we obtain 6 chirally asymmetric 
 vacua and also a chirally symmetric vacuum with $\phi = \mu_{ij} = b = 0$.

 Note that, for $\lambda =0$, we can forget about the moduli $B$ altogether
 and the effective lagrangian has exactly the same form as for the
 supersymmetric QCD with the completely broken $SU(4)$ gauge group.
 The theory has BPS domain walls connecting the chirally symmetric vacuum with
 chirally asymmetric ones and also the walls interpolating between different
 asymmetric vacua whose dynamics depends on the value of mass $m$ in
 a nontrivial way \cite{SUN}. The dynamics of the domain walls with nonzero
 $\lambda$ is yet to be studied.

\section{Discussion.}
Our main result that modifying the tree--level superpotential by adding 
Yukawa terms can bring about new vacuum states coming from infinity of the
moduli space is rather natural. One can  remind in this respect that
the number of
vacua in the supersymmetric Wess--Zumino model is determined by the form
of the superpotential. For a polynomial superpotential, the number of vacua
is $n-1$ where $n$ is the highest power in the polynomial. If we add to the
superpotential the term $X^{n+1}$ with a small coefficient, a new vacua
at large values of $|x_{\rm vac}|$ appears and the qualitative picture is
exactly the same as in Fig. 2.

 Bearing this in mind, it is even surprising that we managed to find only
 one clean example based on the $G_2$ gauge group where extra vacuum
 states appear. $O(7)$ example was a lame one: one cannot really consider
 a non-renormalizable theory with  quartic term in the superpotential.
 To keep a theory renormalizable, we are allowed to add only cubic terms
 with  dimensionless couplings.
 A distiguishing feature of $G_2$ is the presence of a triple group
 invariant $f^{\alpha\beta\gamma} p_\alpha q_\beta r_\gamma$
 made out of fundamental representations. Such an invariant is
 absent for conventional groups (an exception is $SU(3)$, but to be able
 to invoke a cubic term there we must have at least 3 flavours in our disposal
 in which case an instanton-induced superpotential is not generated and the
 physics of the model is different \cite{brmog}). May be one could construct
 further renormalizable examples of the theories where the number
 of asymmetric vacua is not given by the Dynkin index counting by
 considering more complicated 
 representations, playing around with higher exceptional groups where triple
 invariants are typically present and/or relaxing the condition that the gauge
 group is broken completely.

 Note that our finding does not modify the conclusion that the number
of chirally asymmetric vacua
 in a {\it pure} supersymmetric Yang--Mills theory is always $T(G)$. The
 latter is obtained from supersymmetric QCD by tending the masses of the
 matter fields to infinity. If doing that in $G_2$ theory with keeping
 $\lambda$ fixed (we have to keep $\lambda$ small enough not to run in
the Landau pole problem), 
 two extra vacua run away to infinity and decouple.

 The main lesson to be learned from our result is that the number of vacua
 in a supersymmeric theory is not necessarily related to any discrete symmetry
 of the theory which is spontaneously broken:  there is {\it no} trace of
 a $Z_6$ -- symmetry in our model.
 Earlier, we constructed a two--dimensional example with several vacuum states
 whose presence did not follow from symmetry considerations \cite{QCD2}. This
  example was  not even supersymmetric: it was just  QCD$_2$
involving adjoint rather than fundamental fermions and based on the 
  $SU(N)$ gauge group with even $N \geq 4$ . The presence of the
 chirally symmetric vacuum state in 
$4D$  supersymmetric QCD is also not dictated by any symmetry considerations.

 In the 2--dimensional case, we did not manage, however,  to construct these 
 degenerate vacuum states {\it explicitly} and only presented strong arguments
 in favour of their existence. Likewise, the chirally symmetric phase in 
 SQCD$_4$ is seen only in the framework of the TVY lagrangian whose status 
 is not   completely clear. On the contrary,
the effective lagrangian (\ref{WfullG2}) with the kinetic terms
(\ref{kinmod}) is a true  Born--Oppenheimer lagrangian if the
conditions
 $|m| \ll 1, |m/\lambda^{3/4}| \ll 1$ are fulfilled, and the existence of
 6 chirally asymmetric states in this theory is a {\it theorem}. 

 An interesting question to be studied is whether these extra vacuum states
 are still present in the theory at finite volume. Based on the fact that it 
 is not so in QCD$_2$ \cite{QCD2}, neither it is the case for the chirally 
 symmetric phase \cite{SV}, our guess is that they are {\it not}.
That would imply that, at finite volume (however large it is),  the
 energy of two extra states
we have found is not strictly zero like it is not strictly zero 
for  the chirally symmetric state.  It should tend to zero exponentially
fast in the limit $V \to \infty$.

{\bf Acknowledgments}: \hspace{0.2cm} 
 I acknowledge the illuminating discussions with M. Ol'shanetsky and
 I. Polyubin.
This work was supported in part  by the RFBR--INTAS grants 93--0283 and
 94--2851, by the RFFI grant 97--02--16131, 
by the U.S. Civilian Research and Development Foundation under award 
\# RP2--132, and by the Schweizerishcher National Fonds grant \# 7SUPJ048716.

\end{document}